\documentclass[aps,nofootinbib,11pt]{revtex4-1}
\usepackage[letterpaper,hmargin=1in,vmargin=1in]{geometry}

\usepackage{graphicx,epstopdf,amsmath,amsfonts,amssymb,amsxtra,color}

\usepackage{multirow,array}
\usepackage{cancel,bbold,extarrows}
\usepackage{hyperref}

\begin{document}
 
\title{A MASSLESS FIREWALL}

\author{Werner Israel}
\email{israel@uvic.ca}
\affiliation{Department of Physics and Astronomy, University of Victoria, 
     Victoria, B.C., Canada, V8W 2Y2}

\date{April 2014}

\begin{abstract}
Several anomalies of black hole thermodynamics are resolved if one accepts hints from semi classical theory that in regions of strong gravity the energy density of the vacuum can become substantially negative. This leads to a picture of the horizon as a hot massless shell. As viewed in the local (Boulware) ground state, the apparent horizon is the repository of a large store of energy, and this energy is thermal. But, as a source of gravity, its mass is near zero. A simple (1+1)-dimensional model provides a concrete realization of this picture.

\end{abstract}
\maketitle

\section{Introduction}\label{intro}

Quantum fields around a massive star have the property that the energy density of their ground (i.e., Boulware) states is depressed below zero by gravity \cite{RecRev}. The depression grows as one moves inwards and becomes critical for a star contracting toward its gravitational radius. What happens next has not been studied in detail, but this impending singularity appears to be the trigger for releasing a steady flow of Hawking radiation once a horizon has been formed. The outflow fills up the constantly reforming Boulware energy depression at the horizon and so preserves regularity of the geometry there.

The initial Boulware state was cold but the Hawking radiation has a thermal spectrum. So the new (Unruh) state of the field is hot, though its energy density at the horizon is near zero.

It is common to assume implicitly that at this point the ground state of the field has evolved unitarily from the Boulware state to the Unruh state, and to track the subsequent evolution taking the Unruh state as ground state. This assumption gives a satisfactory accounting of energies (and hence of gravitational interactions): the Hawking excitation energy balances the Boulware deficit. But the thermality of the radiation remains uncompensated and overlooked. This leads to a skewed description of thermal effects.

In reality, it appears that the Unruh state is not the unitarily-evolved ground state. It is rather a thermal excitation of that ground state, which remains the energetically depressed Boulware state. Near the horizon the Boulware depression is offset by Hawking radiation which brings up the energy density there to the Unruh near-zero level.

For purposes of a semi-classical theory there are obvious advantages to working with expectation values in the actual (Unruh) state rather than the Boulware ground state. The downside is a distorted picture of thermal effects, since the thermality (though not the stress-energy) of the Hawking radiation has been passed over.

We must then be prepared for the (retrospectively obvious) consequence that, in this new (Unruh) gauge, \emph{the large thermal energies near the horizon have only small gravitational and inertial effects}, and contribute little to the stress-energy on the right-hand side of Einstein's equations. Near the horizon we have heat without mass.

This unveils a picture of the horizon as a nearly massless shell which nevertheless contains a large store of thermal energy---a ``firewall''.

Remarkably, such a possibility was adumbrated recently on quite different grounds by Almheiri, Marolf, Polchinski and Sully \cite{Polchinski}. They advocated a firewall as the ``most conservative'' solution to a paradox that arises when one examines correlations of late modes of Hawking radiation with their internal partners and with the early modes.

The firewall has earlier antecedents. 't Hooft's ``brick wall'' model \cite{tHooft,Israel} for black hole entropy was a proposal to explain the Bekenstein-Hawking area law by actually localizing the entropy on the horizon. It now emerges that the brick wall is a firewall. Even earlier, there are connections with the ``membrane paradigm'' of Thorne and associates \cite{membrane_paradigm}.

\section{Reduced Spherical Einstein Theory}

For a spherical metric in the general form
\begin{equation}
\operatorname{d}\!s^2=g_{ab}\operatorname{d}\!x^a\operatorname{d}\!x^b+r^2(x^a)\operatorname{d}\!\Omega^2
\end{equation}
($a,b=0,1$), the 4-dimensional Einstein-Hilbert Lagrangian reduces to
\begin{equation}\label{Lagrn}
L=\frac{1}{4}r^2R+\frac{1}{2}(\nabla r)^2+L_{\mathrm{mat}},
\end{equation}
where $R$ is the 2-dimensional curvature scalar.

We shall adopt this same Lagrangian to define a ``spherical'' Einstein theory in (1+1) dimensions, with $r(x^a)$ now an auxiliary scalar (``dilaton'') field.

Define the scalar fields $f,m$ by
\begin{equation}
f=(\nabla r)^2=1-2m/r
\end{equation}
Then \eqref{Lagrn} leads to the field equations \cite{poisson_israel}
\begin{align}
\partial_a m&=(T_a^b-\delta_a^bT_d^d)\partial_br\label{field_eq1}\\
r_{;ab}&=\bigl(\Box r-\frac{m}{r^2}\bigr)g_{ab}-\frac{1}{r}T_{ab}\label{field_eq2},
\end{align}
which imply the conservation laws
\begin{equation}
T^b_{a;b}=0.
\end{equation}

For access to the future horizon it is useful to work with an advanced time co-ordinate $v$. A general 2-metric then takes the form
\begin{equation}\label{metric_v}
\operatorname{d}\!s^2=2e^\psi\operatorname{d}\!v\operatorname{d}\!r-fe^{2\psi}\operatorname{d}\!v^2
\end{equation}
and the field equations \eqref{field_eq1}, \eqref{field_eq2} become
\begin{equation}\label{field_eq3}
m_r=-T_v^v,\qquad m_v=T_v^r,\qquad \psi_r=\frac{1}{r}T_{rr}.
\end{equation}
(Subscripts on $m$ and $\psi$ indicate partial derivatives.)

The curvature scalar for metric \eqref{metric_v} is
\begin{equation}\label{Rscalar}
R=-2e^{-\psi}(\alpha+\psi_v)_r,
\end{equation}
where $\alpha=\frac{1}{2}e^{-\psi}(fe^{2\psi})_r$ is related to the local surface gravity
\begin{equation}
\kappa\equiv(-g_{00})^{\frac{1}{2}}a=\alpha-\frac{1}{2}\partial_v(\ln f)
\end{equation}
(=redshifted acceleration $a$ of a static observer $r=\mathrm{const.}$).

\section{Hawking Radiation}

As source $T_a^b$ we consider a quantized massless scalar field propagating on this classical background. Expectation values of all components $T_a^b$ can be obtained from the conservation laws and the 2-dimensional conformal trace anomaly \cite{RecRev}
\begin{equation}\label{traceT}
T_a^a=\tilde{h}R,\qquad\tilde{h}\equiv\hbar/24\pi.
\end{equation}

By manipulation of the conservation laws and use of \eqref{traceT} one arrives at
\begin{equation}\label{conserv_r}
\partial_r\{fe^{2\psi}T_r^r+2e^{\psi}T_v^r+\tilde{h}(\kappa^2-2\partial_v\kappa)\}=0
\end{equation}
(details in the Appendix). This provides a convenient first integral of the conservation laws.

It is intuitively helpful to put physical flesh onto these equations by introducing an orthonormal basis ($s^a,t^a$) anchored to the curves $r=\mathrm{const.}$. The unit tangent to these curves is labelled $t^a$ in the exterior domain $f>0$ where they are timelike, and $s^a$ in the black hole interior $f<0$. We also define the outgoing and ingoing lightlike vectors
\begin{equation}
l^a=t^a+s^a,\qquad n^a=t^a-s^a.
\end{equation}

The stress tensor $T^{ab}$ for any state (but we have in mind the Unruh state) can then be decomposed into a (Boulware-like) fluid part and an outflux $F$:
\begin{equation}\label{decomp_B}
T^{ab}=T_\mathrm{B}^{ab}+Fl^al^b
\end{equation}
where
\begin{equation}
T_\mathrm{B}^{ab}=P_\mathrm{B}s^as^b+\rho_\mathrm{B}t^at^b
\end{equation}
This is useful in the exterior, where the Unruh boundary conditions require $(P_\mathrm{B},\rho_\mathrm{B})\rightarrow0$ at past infinity and we want to isolate the Hawking flux $F$.

Alternatively and equivalently, we can decompose into a (Hartle-Hawking-like) fluid part and an oppositely-signed influx:
\begin{equation}\label{decomp_H}
T^{ab}=T_\mathrm{H}^{ab}-Fn^an^b
\end{equation}
with
\begin{equation}\label{Prho_HB}
P_\mathrm{H}=P_\mathrm{B}+2F,\qquad\rho_\mathrm{H}=\rho_\mathrm{B}+2F.
\end{equation}
This provides a clearer description of the interior by exposing the infalling flux of negative energy from the horizon that accompanies the Hawking outflux.

For the Unruh stress tensor in the decomposition \eqref{decomp_B}, equation \eqref{conserv_r} reduces to
\begin{equation}\label{conserv_r1}
\partial_r\Bigl[\tilde{h}(\kappa^2-2\partial_v\kappa)+\lvert f\rvert e^{2\psi}\begin{cases}
P_\mathrm{B}&(f>0)\\
\rho_\mathrm{B}&(f<0)
\end{cases}\;\Bigr]=0
\end{equation}
(the flux term $F$ drops out). Hence the expression in square brackets can be an arbitrary function of $v$, which must however vanish to satisfy the Unruh boundary condition that the spacetime is empty at past lightlike infinity. Hence \eqref{conserv_r1} integrates to
\begin{equation}\label{conserv_r1_int}
\left.\begin{aligned}
P_\mathrm{B}&\quad(f>0)\\
\rho_\mathrm{B}&\quad(f<0)
\end{aligned}\right\}
=-\tilde{h}\frac{\kappa^2-2\partial_v\kappa}{\lvert f\rvert e^{2\psi}}.
\end{equation}
The singularity this produces in $T_\mathrm{B}^{ab}$ on the apparent horizon $f=0$ (and hence threatens in the Unruh stresses) must be offset by the flux term in \eqref{decomp_B}.

Equation \eqref{field_eq3} for $\psi$ becomes
\begin{equation}\label{field_eq3_psi}
rf\psi_r=\rho_\mathrm{B}+P_\mathrm{B}+4F
\end{equation}
and the trace anomaly \eqref{traceT} gives
\begin{equation}
P_\mathrm{B}-\rho_\mathrm{B}=\tilde{h}R.
\end{equation}

From \eqref{field_eq3_psi}, one sees that regularity of $\psi$ at $f=0$ (hence regularity of $R$ by \eqref{Rscalar}) requires the Hawking flux there to be given by
\begin{equation}\label{Hawking_flux}
\lvert f\rvert e^{2\psi}F=\frac{1}{2}\tilde{h}(\kappa_0^2-2\partial_v\kappa_0)\quad(f\rightarrow0),
\end{equation}
where $\kappa_0(v)$ is the surface gravity $\kappa$ at the apparent horizon.

For slow evaporation, the expression on the left of \eqref{Hawking_flux} is nearly conserved, as can be seen from
\begin{equation}
\frac{\operatorname{d}}{\operatorname{d}r}(fe^{\psi}F)=\frac{4}{r}e^{\psi}F^2,
\end{equation}
which follows from the conservation laws ($\operatorname{d}\!/\!\operatorname{d}r$ is the rate of change along outgoing light rays). Thus the right-hand side is also nearly the Hawking flux at infinity:
\begin{equation}
F(r\rightarrow\infty)\approx\frac{1}{2}\tilde{h}\kappa_0^2
\end{equation}
This is the standard expression for the Hawking flux in (1+1) dimensions.

We have thus arrived at a picture of the apparent horizon as the source of two opposing and oppositely-signed streams of thermal energy. The matter content of the horizon itself is most directly appreciated from $T_\mathrm{H}^{ab}$ in the decomposition \eqref{decomp_H}, because here the added flux term is transverse, hence not blueshifted. Equations \eqref{Prho_HB}, \eqref{conserv_r1_int} and \eqref{Hawking_flux} show that $T_\mathrm{H}^{ab}$ vanishes in highest order $f^{-1}$ and hence is at most of order $\hbar/m^2$ when $f=0$. The apparent horizon is virtually empty---but hot!

\section{Summary and Conclusions}

With the recognition that the apparent horizon of a black hole, far from being inactive, is the seat of a very hot, yet massless shell---a ``firewall'' \cite{Polchinski}---a number of curious features of black hole thermodynamics fall naturally into place. The reason why the hole's thermal parameters are linked to its surface properties---entropy to surface area, temperature to surface gravity---is that they \emph{are} just surface properties. Hawking radiation is just thermal radiation from a hot surface.

Most intriguingly, this may hold the key to a conservative solution of the puzzle of information loss. Dropping information into a black hole is, after all, not so different from throwing an encyclopaedia in the fire. All information can, in principle, be recovered by collecting and reassembling the smoke and ashes.

\appendix*

\section{Derivation of (\ref{conserv_r})}

If the two conservation laws $T^b_{a;b}=0$ are written out, and terms in $\partial_vf$ are eliminated between them, the result can be cast in the form
\begin{equation*}
fe^{\psi}\partial_rT_r^r+2e^{-\psi}\partial_r(e^{\psi}T_v^r)+(T_r^r-\frac{1}{2}T_a^a)e^{-\psi}\partial_r(fe^{2\psi})+\partial_vT_a^a=0,
\end{equation*}
Inserting the trace anomaly \eqref{traceT} for $T_a^a$ and the expression \eqref{Rscalar} for $R$ into this equation leads, after some manipulations, to \eqref{conserv_r}.

\section*{Acknowledgements}

Thanks to Xun Wang for help with the manuscript, and to him and Adam Ritz for useful comments.

\end{document}